\documentclass[11pt]{article}
\usepackage[dvips]{graphicx}
\begin{document}
\begin{center}
{\LARGE\bf A Comparison of Quintessence and Nonlinear Born-Infeld
Scalar Field Using Gold Supernova data  } \vskip 0.15 in
$^\dag$Wei Fang$^1$, $^\ddag$H.Q.Lu$^1$, B.Li$^2$, K.F Zhang$^1$\\
$^1$Department~of~Physics,~Shanghai~University,\\
~Shanghai,~200444,~P.R.China\\
$^2$Department~of~Physics,~The~Chinese~University~of~Hong~Kong,\\
~Hong ~Kong~ SAR, China
\footnotetext{$\dag$Fangwei@graduate.shu.edu.cn}
\footnotetext{$\ddag$ Alberthq$_-$lu@staff.shu.edu.cn}

 \vskip 0.5
in \centerline{\textbf{Abstract}}
\end{center}
We study the Non-Linear Born-Infeld( NLBI) scalar field model and
quintessence model with two different potentials( $V(\phi)=-s\phi$
and $\frac{1}{2}m^2\phi^2$). We investigate the differences
between those two models. We explore the equation of state
parameter w and the evolution of scale factor $a(t)$ in both NLBI
scalar field and quintessence model. The present age of universe
and the transition redshift are also obtained. We use the Gold
dataset of 157 SN-Ia to constrain the parameters of the two
models. All the results show that NLBI model is slightly superior
to quintessence model. \vskip 0.2 in
\begin{minipage}{5.5in}
{ \hspace*{15pt}\small \\
{\bf Keywords:}Dark energy;~Quintessence Model;~NLBI Model.\\
 {\bf PACS:}98.80.Cq}
\end{minipage}

\section{Introduction}
 \hspace*{15pt} According to analyses of
 CMB[1]+SN-Ia[2]+HST[3]+LSS[4]
 data, there are strong evidences to indicate that our universe has
 recently entered a phase of accelerating expansion and that the
 universe is flat. It implies that if this is not a signal
 of modifying the standard theory of gravity, that may be a evidence
 of existing a "exotic homogeneous matter" with negative pressure
 termed "dark energy"(DE)[5]. An equation of state parameter (w$=p/\rho$)
 is usually used to describe this energy component. The value of
 w is required to less that $-\frac{1}{3}$ for a accelerating
 expansion universe. Due to the existence of ordinary matter and
 radiation, we actually need a more negative value of w to drive
 accelerating expansion of the universe. Analysis shows w can lie in
 the rang of $-1.32<$w$<-0.82$ with the 2-$\sigma$ confidence
 levels[6]. Up to now a cosmological constant with w$=-1$ is in good
 agreement with all the data. However, the result
 doesn't rule out the scalar field and phantom field models as the
 dark energy candidate. For the $\Lambda$CDM model[7], there are two
 boring problems: the fine-tuning problem and cosmic coincidence
 problem[8]. Though same problems remain in other models, it
 can be alleviated in some models[9].
 \par The scalar field appeared in cosmology is not the first
 time. As early as more than twenty years ago, theorist normally
 consider a homogenous scalar field $\phi$ in an inflationary universe[10]. A
 scalar field $\phi$ slowly evolving down its potential $V(\phi)$
 can drive both inflation and late-time accelerating expansion. The
 standard quintessence scenario[11](a canonical scalar field described
 by a lagrangian $L=\frac{1}{2}{\dot\phi}^2-V(\phi)$) is a simplest scalar field.
 The universes they predict will have
 significant differences with respect to different potentials.
 Another more complicated scalar field is K-essence. The idea of
 K-essence was firstly introduced as a possible model for inflation[12].
 Later it was noted that K-essence was introduced as a possible models
 for dark energy[13]. K-essence can be defined as any scalar field
 with non canonical kinetic terms. Its lagrangian usually takes
 the form $L_A=V(\phi)F(X)$. A more general form of lagrangian for
 K-essence[14] is $L_B=f(\phi)g(X)-V(\phi)$, where $\phi$ is the
 scalar field, and
 $X=\frac{1}{2}\nabla_{\mu}\phi\nabla^{\mu}\phi$.
 \par The Born-Infeld[15] theory has been considered widely in string
 theory and cosmology[16]. There are two Born-Infeld type scalar
 field models. The first is rolling tachyon field[17] with the lagrangian
 form $L_{tach}=-V(\phi)\sqrt{1-{\dot\phi}^2}$, which can be
 classified as $L_A$. Its interesting features have been widely
 studied[18]. Anther one is the Nonlinear Born-Infeld(NLBI) scalar field
 theory with the lagrangian form
 $L_{NLBI}=\frac{1}{\eta}[1-\sqrt{1-\eta{\dot\phi}^2}]-V(\phi)$(noted,
 when field velocity $\dot\phi
 \rightarrow0$,$L_{NLBI}=L_{quin}=L=\frac{1}{2}{\dot\phi}^2-V(\phi)$ by Taylor
 expansion). Obviously it can be regarded as one form of $L_B$. W.Heisenberg proposed
 this Lagrangian density in order to describe the process of meson
 multiple production connected with strong field regime[19]. H.P.de Oliveira qualitatively studied the solutions of
 a three-dimensional dynamical system describing the static and spherically symmetric solutions for this NLBI scalar field [20].
  The dark energy model with this NLBI scalar field was recently suggested by H.Q.Lu[21]. However, comparing
 with tachyon field, the role of NLBI scalar field in cosmology is far
 beyond study.
 \par One direct thought motivated us to consider the NLBI field
is, besides the attractive properties[19,22], being able to study
the role of nonlinearities in the matter fields. Furthermore, we
will be able to see whether it can provide more interesting
 physical results than those generated by ordinary quintessence
 model.
 \par It is important to choose the potential
 $V(\phi)$ for the scalar field. In many cases, we use a potential
 predicted by particle physics. However, we do not really know
 which theory of particle physics best describes the universe, we should keep an open mind as to the form of $V(\phi)$.
 One ideal approach is to consider some simple and possible
 forms of potential, explore the cosmological behavior in detail using the qualitative theory of dynamical systems, and
 then study the evolution of universe in more detail to
 quantitatively fit with current observation data. We choose two
 simple potentials $-s\phi$ and $\frac{1}{2}m^2\phi^2$ for further
 study in this paper. Constraints on the linear potential $-s\phi$ in quintessence  and
 phantom models from recent supernova data have been argued in[23]. It has been argued[24] that such a
 potential is favored by anthropic principle considerations and
 can provide a potential solution to the cosmic coincidence
 problem. The square potential
 $\frac{1}{2}m^2\phi^2$ has been considered in a chaotic inflationary universe[25].
 \par In this paper, we will focus on the differences between NLBI
 scalar field and linear scalar field(quintessence), explore the
 cosmological scenario in detail and compare them with
 SN-Ia Gold data. This paper is organized as follows: In
 section 2, we simply review the theoretic model with the two scalar fields and derive their cosmological evolution with two different
 potentials. In section 3, we fit the Hubble parameter to the
 SN-Ia Gold data, obtain constraint for the potential
 parameters and compare the NLBI scalar field with the linear scalar field.
 Section 4 is conclusion and discussion.

\section{Theoretic Models of Quintessence and NLBI Scalar Field  }\hspace*{15pt}
  \par We consider a spatially homogeneous scalar field in a flat universe with Robertson-Walker metric
  $ds^2=dt^2-a^2(t)(dx^2+dy^2+dz^2)$, here we have neglect the
  relativistic radiation component and assume only presence of
  non-relativistic matter component  and scalar field.\\
  \textbf{1.Linear Scalar Field}:
\begin{equation}L_{quin}=\frac{1}{2}{\dot\phi}^2+V(\phi)\end{equation}
\begin{equation}\frac{\dot a}{a}=\frac{1}{\sqrt3Mp}[\frac{1}{2}{\dot\phi}^2-V(\phi)+\rho_0\frac{a^3_0}{a^3}]^{\frac{1}{2}}\end{equation}
\begin{equation}\ddot\phi+3H\dot\phi+\frac{dV(\phi)}{d\phi}=0\end{equation}
\begin{equation}w=\frac{\frac{1}{2}{\dot\phi}^2-V(\phi)}{\frac{1}{2}{\dot\phi}^2+V(\phi)}\end{equation}
\begin{equation}c^2_s=1\end{equation}
 \textbf{2.NLBI Scalar Field}:
\begin{equation}P_{NLBI}=L_{NLBI}=\frac{1}{\eta}[1-\sqrt{1-\eta{\dot\phi}^2}]-V(\phi)\end{equation}
\begin{equation}\rho_{NLBI}=\frac{1}{\eta\sqrt{1-\eta{\dot\phi}^2}}-\frac{1}{\eta}+V(\phi)\end{equation}
\begin{equation}\frac{\dot a}{a}=\frac{1}{\sqrt3Mp}[\frac{1}{\eta\sqrt{1-\eta{\dot\phi}^2}}-\frac{1}{\eta}+V(\phi)+\rho_0\frac{a^3_0}{a^3}]^{\frac{1}{2}}\end{equation}
\begin{equation}\ddot\phi+(1-\eta{\dot\phi}^2)[3H\dot\phi+\frac{dV(\phi)}{d\phi}(1-\eta{\dot\phi}^2)^{\frac{1}{2}}]=0\end{equation}
\begin{equation}w=\frac{1-\sqrt{1-\eta{\dot\phi}^2}-\eta V(\phi)}{\frac{1}{\sqrt{1-\eta{\dot\phi}^2}}-1+\eta V(\phi)}\end{equation}
\begin{equation}c^2_s=1-\eta{\dot\phi}^2\end{equation}
We can see from above equations, both the field equation(Eq.(9))
and sound speed of NLBI scalar field(Eq.(11)) will recover to
quintessence model(Eqs.(3,5) if $\dot\phi\rightarrow0$. This is
consistent with above description of NLBI lagrangian. No matter
for quintessence or NLBI scalar field it is worth noting that
potential($V(\phi)$) term can not determine whether w crosses the
phantom divide line(PDL),it can not directly determine the value
of sound speed too. Additionally, When the potential
$V(\phi)=\frac{1}{\eta}$, Eq.(6) actually describes the lagrangian
of Chaplygin gas[26]
\par One apparent effect of DE is through its impact on the
expansion rate, which is determined by the equation of state of
the DE component. It gives us an opportunity to explore the
information of DE by measurements of the relationship between
luminosity distance and redshift $z$. Since quintessence and NLBI
scalar field have different evolution  of the equation of state w,
It provides us a possibility to distinguish between quintessence
and NLBI scalar field by comparing with SN-Ia data and future SNAP
data. Another effect of DE is directly through the perturbation to
affect the CMB. This feature inform us another opportunity to test
DE. It is by looking at the anisotropy in the temperature and
polarization of CMB, which have been measured by WMAP and will be
PLANCK mission. The speed of sound of DE mainly affects the CMB
spectrum at largest scalars via the late-time Integrated Sachs
Wolfe effect[27]. Since $c^2_s$ determines how fast fluctuations
dissipate, a lower sound speed increases the phase space of models
which are Jeans unstable[28]. Hence the different value of sound
speed predicted by quintessence model and NLBI scalar field
model($c^2_s=1$ for quintessence and $c^2_s=1-\eta{\dot\phi}^2$
for NLBI scalar field) makes the spectra of anisotropies of the
two models slightly different because of the different Jeans
scales in the DE sector. So we have another possibility to
distinguish between quintessence and NLBI scalar field  by
analysing the WMAP data and other observational data. In this
paper, we focus our
interest on SN-Ia measurement and remain the latter possibility for further study.\\
\section{The Evolution of Universe of Quintessence and NLBI scalar Field with Two Special Potentials } \hspace*{15 pt}
\par\textbf{A}. $V(\phi)=-s\phi$
 \par By setting $H_0t\rightarrow
 t$,$\phi\rightarrow\sqrt3 M_p\phi$,$s\rightarrow\sqrt3H^2_0 s$, the
 evolution equation of scale factor and the scalar field equation
 for quintessence model can be written in rescaled form as
\begin{equation}\frac{\dot a}{a}=[\frac{1}{2}{\dot\phi}^2-s\phi+\frac{\Omega_{om}}{a^3}]^{\frac{1}{2}}\end{equation}
\begin{equation}\ddot\phi+3\frac{\dot a}{a}\dot\phi-s=0\end{equation}
By the same transformation, the corresponding equations for NLBI
model are
\begin{equation}\frac{\dot a}{a}=[\frac{1}{\sqrt{1-\eta{\dot\phi}^2}}-1-s\phi+\frac{\Omega_{om}}{a^3}]^{\frac{1}{2}}\end{equation}
\begin{equation}\ddot\phi+(1-{\dot\phi}^2)[3\frac{\dot
a}{a}\dot\phi+s(1-{\dot\phi}^2)^{\frac{1}{2}}]=0\end{equation}
where we set $\eta=1/\rho_c=1/3M^2_p H^2_0$. We follow Ref[23] and
solve the system numerically using the following initial
conditions($ t\rightarrow t_i\simeq0$):
$a(t_i)=(\frac{9\Omega_{0m}}{4})^{1/3}t^{2/3}_i$,
$\dot\phi(t_i)=0$, $\phi(t_i)=\phi_i$. The numerical program[29]
is modified to be suited for us. The value of $\phi_i$ is chosen
for each $s$ such that
$\Omega_{0\phi}=\frac{1}{\sqrt{1-{\dot\phi(t_0)}^2}}-1-V(\phi(t_0))=1-\Omega_{0m}$
at the present time $t_0$(for quintessence model
$\Omega_{0\phi(t_0)}=\frac{1}{2}{\dot\phi}^2(t_0)+V(\phi(t_0))$ ).
We set $a(t_0)=1$ and assume a prior of $\Omega_{0m}=0.3$. From
Eq.(3), we can easily obtain
\begin{equation}\rho_{quin}=-3\int H{\dot\phi}^2dt+const\end{equation}
We can also obtain a similar equation for NLBI scalar field:
\begin{equation}\rho_{NLBI}=-3\int H\frac{{\dot\phi}^2}{\sqrt{1-\eta{\dot\phi}^2}}dt+const\end{equation}
These two equations inform that only in a static universe, the
energy density of scalar field is unchanged. In a
expansive(contractive) universe $H>0(H<0)$, the energy density of
scalar field rolls down(up) the potential for both quintessence
and NLBI scalar field model. The numerical results are plotted in
fig1-4.
\begin{center}\vspace{0.5cm}
\includegraphics[width=10cm]{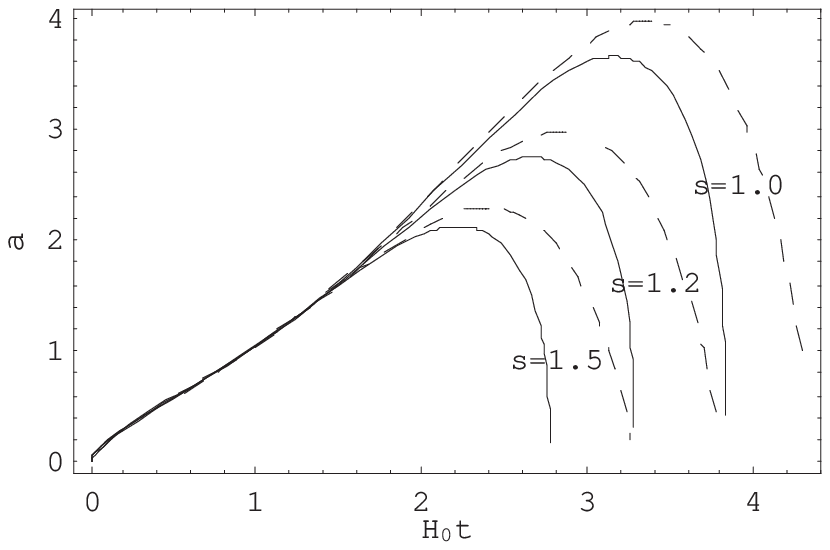} \end{center} \hfill
~\begin{minipage}{5.5in}~Fig1.The evolution of scalar factor for
quintessence field and NLBI scalar field with linear negative
potential. Solid line for quintessence, dot line for NLBI scalar
field.The time that universe begins to contract in NLBI scalar
field is later than that in quintessence models. This feature is
shown clearly in fig4.
\end{minipage}
\begin{center}\vspace{0.5cm}
\includegraphics[width=10cm]{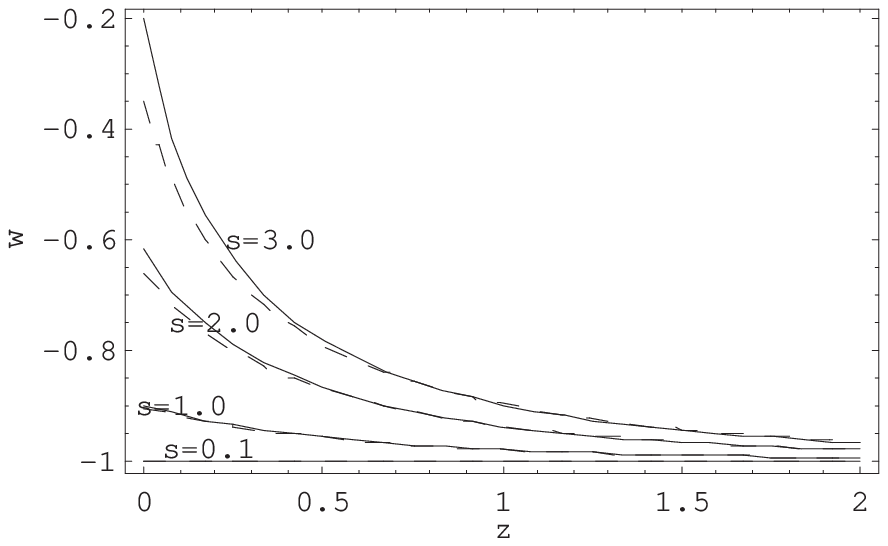} \end{center} \hfill
~\begin{minipage}{5.5in}~Fig2. The evolution of the equation of
state w(z) with respect to redshift $z$. Solid line for
quintessence, dot line for NLBI scalar field. The result shows
that a low value of parameter $s$(therefore a low value of w(z))
is favored by current observation. For the same parameter $s$,
NLBI model has a lower value of equation of state w, but the
difference is not significant when $s$ is close to zero. The
equation of state w can not cross the PLD for both quintessence
and NLBI model.
\end{minipage}
\begin{center}\vspace{0.5cm}
\includegraphics[width=10cm]{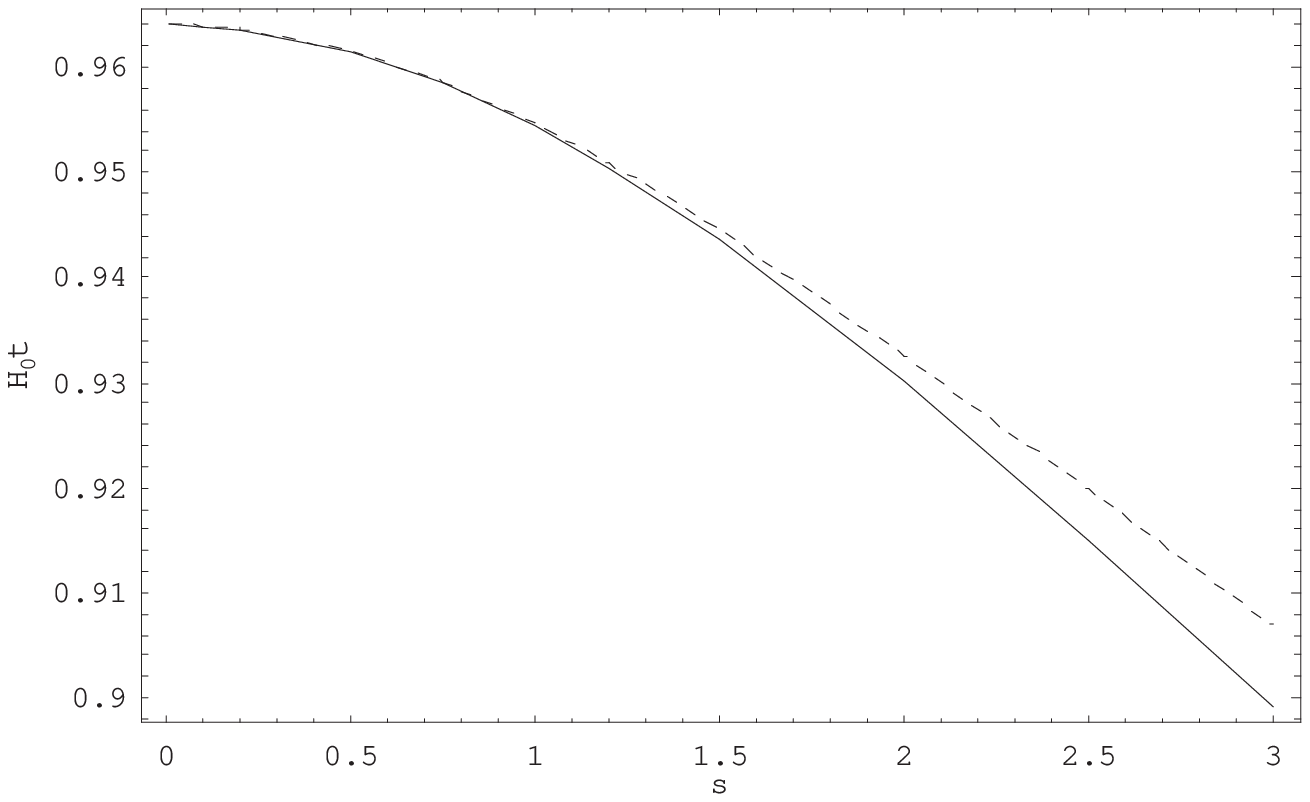} \end{center} \hfill
\begin{minipage}{5.5in}~Fig3. The present age of  universe $H_0t_0$
with different s. Solid line for quintessence, dot line for NLBI
scalar field. The curve has been sampled at $s=0.01, 0.1, 0.2,
0.5, 0.75, 1.0, 1.2, 1.5, 2.0, 2.5, 3.0$ and the corresponding
points have been joined. The result is consistent with observation
to globular cluster and white dwarf[30]. For $s=0.1$, the age of
universe will be 13.77 Gyr for NLBI scalar field model if we take
$H_0=70kms^{-1}Mpc^{-1}$.
\end{minipage}
\begin{center}\vspace{0.5cm}
\includegraphics[width=10cm]{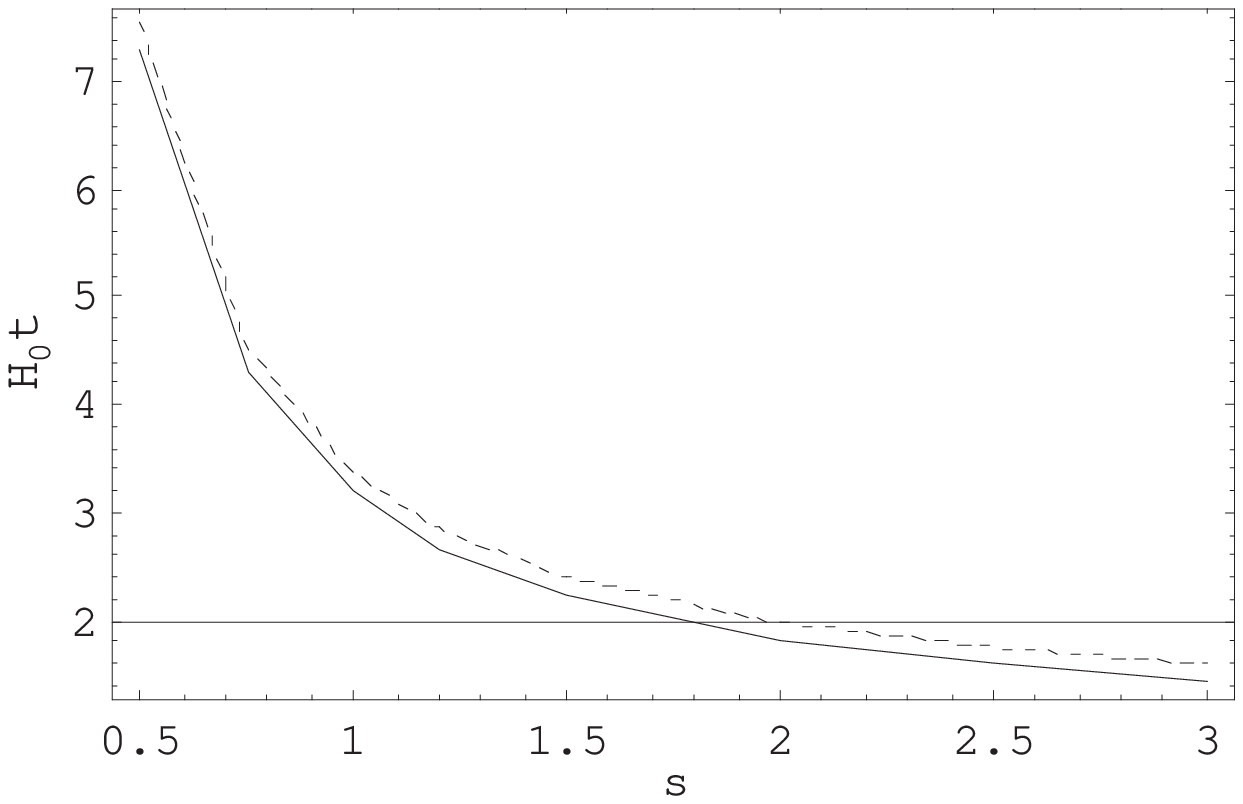} \end{center} \hfill
\begin{minipage}{5.5in}~Fig4. The age of universe $H_0t$ with
different s when scale factor starts to contract. Solid line for
quintessence, dot line for NLBI scalar field. The curve has been
sampled at $s=0.01, 0.1, 0.2, 0.5, 0.75, 1.0, 1.2, 1.5, 2.0, 2.5,
3.0$ and the corresponding points have been joined. It is worth
noting from fig3 and fig4 that the universe with NLBI scalar field
has longer age than the universe with quintessence.
\end{minipage}
\par\textbf{B.} $V(\phi)=\frac{1}{2}m^2\phi^2$\par By setting $H_0t
\rightarrow t, \phi \rightarrow \sqrt 3M_p\phi, m \rightarrow
mH_0$, we can obtain the equations for quintessence:
\begin{equation}\frac{\dot a}{a}=[\frac{1}{2}{\dot\phi}^2-\frac{1}{2}m^2\phi^2+\frac{\Omega}{a^3}]^{\frac{1}{2}}\end{equation}
\begin{equation}\ddot\phi+3\frac{\dot a}{a}\dot\phi-m^2\phi=0\end{equation}
By the same transformation, the corresponding equations for NLBI
model are
\begin{equation}\frac{\dot a}{a}=[\frac{1}{\sqrt{1-\eta{\dot\phi}^2}}-1+\frac{1}{2}m^2\phi^2+\frac{\Omega}{a^3}]^{\frac{1}{2}}\end{equation}
\begin{equation}\ddot\phi+(1-{\dot\phi}^2)[3\frac{\dot
a}{a}\dot\phi+m^2\phi(1-{\dot\phi}^2)^{\frac{1}{2}}]=0\end{equation}
\\
 We
numerically solve the equations with same conditions as negative
linear potential.
\begin{center}\vspace{0.5cm}
\includegraphics[width=10cm]{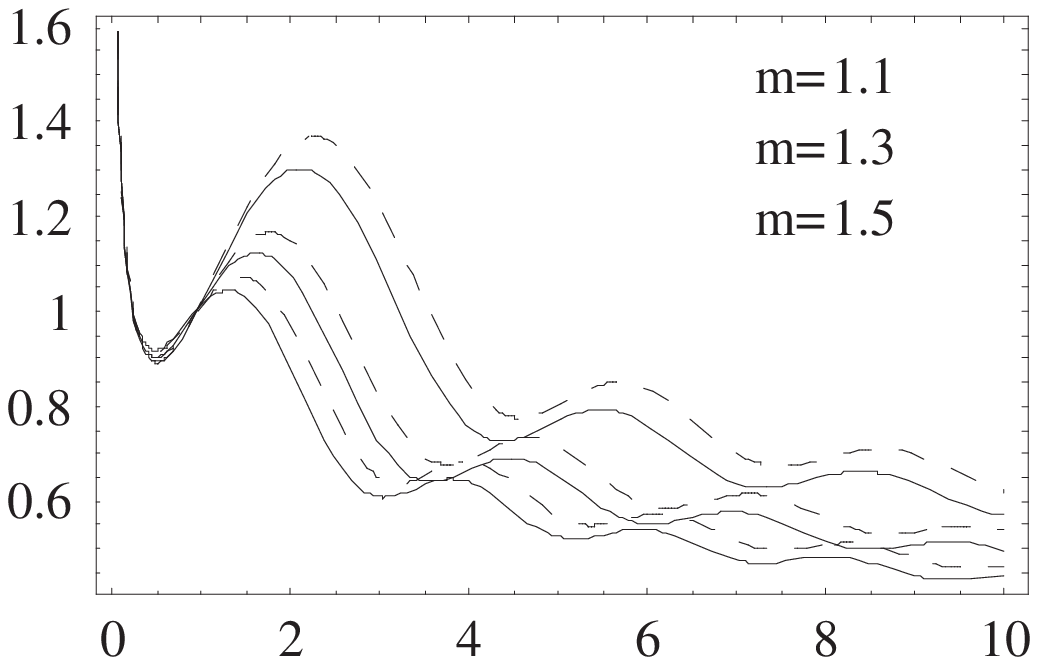} \end{center} \hfill
~\begin{minipage}{5.5in}~Fig5.The evolution of $\dot a$ with
respect to $H_0t$. The horizontal axis represents $H_0t$ and the
vertical axis represents $\dot a$. Solid line for quintessence,
dot line for NLBI scalar quintessence. Unlike negative potential,
the universe with a positive potential will expand for ever. For
square potential, the numerical result indicates that our universe
has entered a phase of accelerated expansion from a phase of
decelerated expansion in the recent past. This is consistent with
observational result[31]. Furthermore, the result also shows that
accelerated expansion and the decelerated expansion will appear by
turns.
\end{minipage}
\begin{center}\vspace{0.5cm}
\includegraphics[width=10cm]{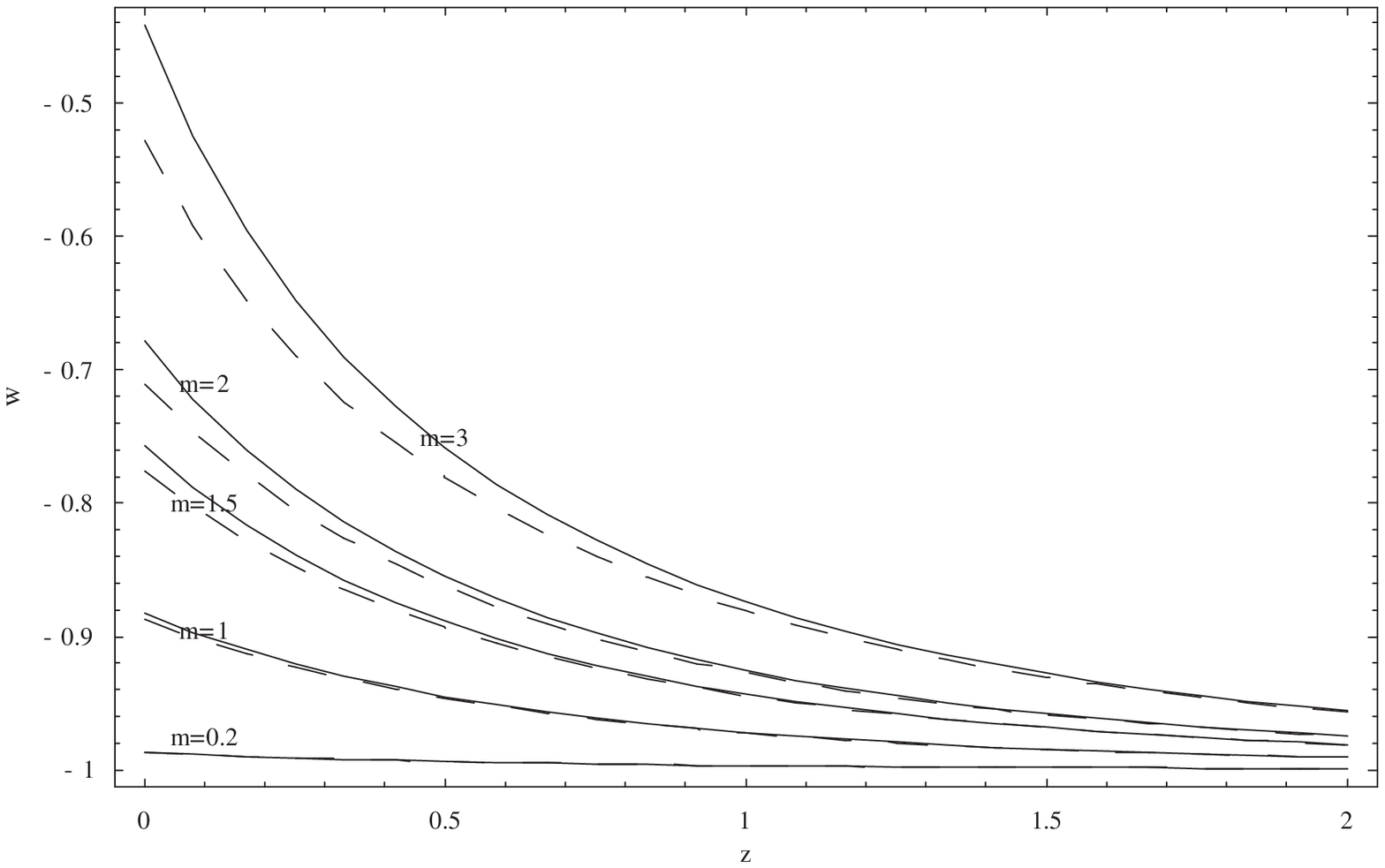} \end{center} \hfill
~\begin{minipage}{5.5in}~Fig6.  The evolution of the equation of
state w(z) with respect to redshift $z$. Solid line for
quintessence, dot line for NLBI scalar quintessence. The result
shown here is similar to that in fig2.
\end{minipage}
\begin{center}\vspace{0.5cm}
\includegraphics[width=10cm]{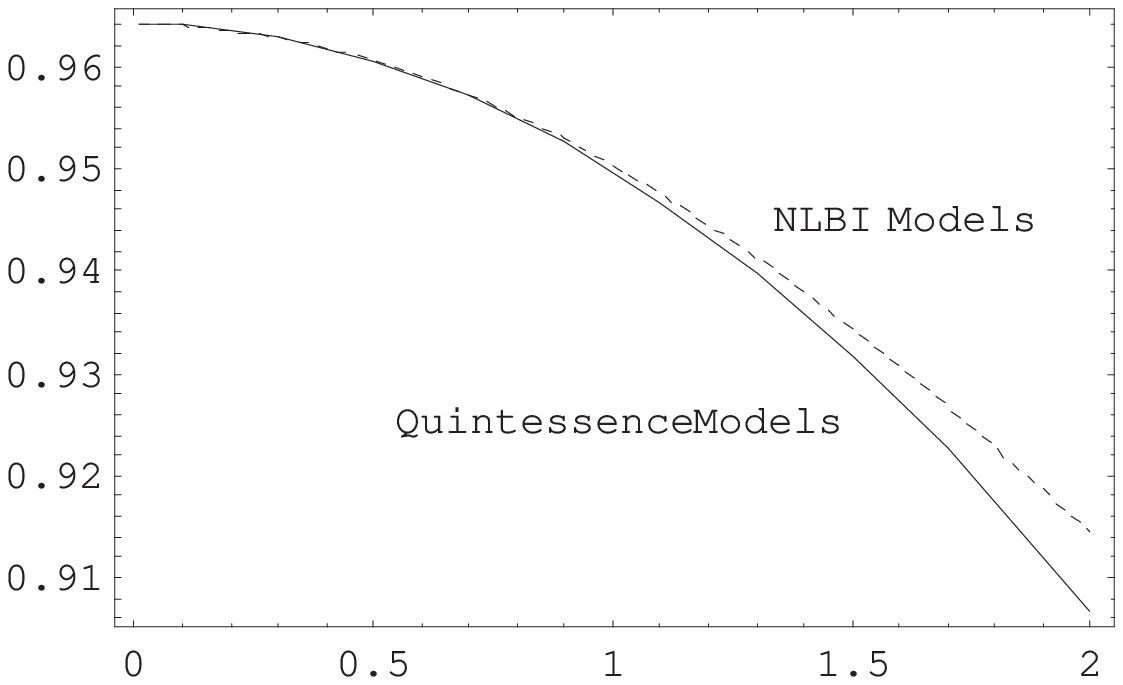} \end{center} \hfill
\begin{minipage}{5.5in}~Fig7. The age of current universe $H_0t$
with different $m$. The horizontal axis represents $m$ and the
vertical axis represents $H_0t$. The result has no significant
difference with that in fig3. The curve has been sampled at
$m=0.01, 0.1, 0.3, 0.5, 0.7, 0.9, 1.1, 1.3, 1.5, 1.7, 2.0$ and the
corresponding points have been jointed.
\end{minipage}
\begin{center}\vspace{0.5cm}
\includegraphics[width=10cm]{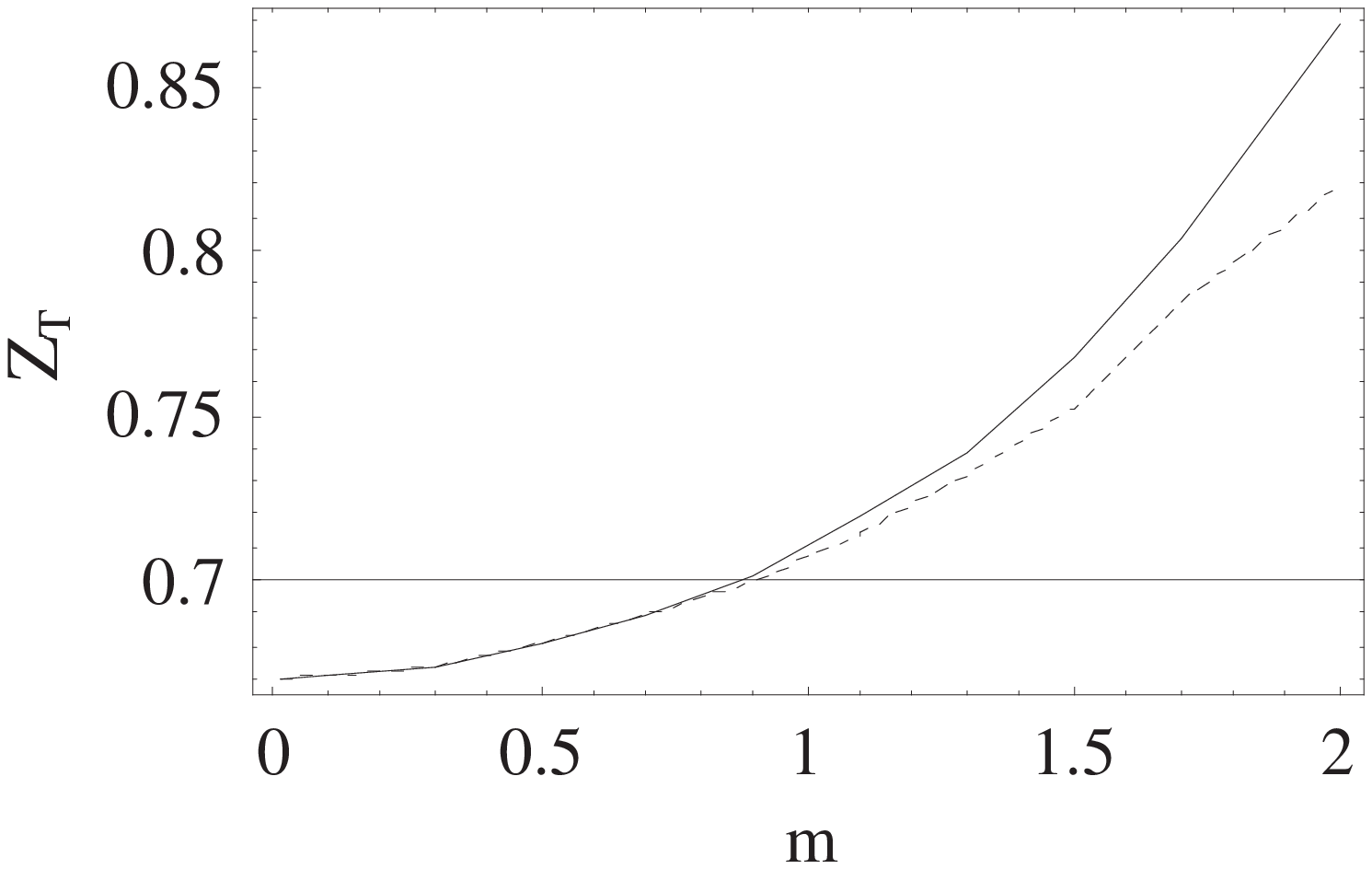} \end{center} \hfill
\begin{minipage}{5.5in}~Fig8. The transition redshift $Z_T$( the redshift that universe
evolves from decelerating expansion to  accelerating expansion)
with different $m$. The result does not conflict with the result
in Ref[31]. The curve has been sampled at $m=0.01, 0.1, 0.3, 0.5,
0.7, 0.9, 1.1, 1.3, 1.5, 1.7, 2.0$ and the corresponding points
have been jointed.
\end{minipage}
\section{Fit to the SN-Ia Gold Data Set}
\par We use the SN-Ia observations to put constraints on
the quintessence and NLBI models. To do this, we need to obtain
the corresponding Hubble parameter $H(z,s\&m)= \frac{\dot
a}{a}(z,s\&m)$ as a function of redshift. The luminosity
distance($d_L$) for a source at redshift $z$ located at radial
coordinate distance $r$ is given by $d_L=(1+z)a_0r$ where $a_0$ is
the present value of the scale factor. Then we can obtain the
corresponding Hubble free luminosity distance in a spatially flat
expanding universe:
\begin{equation}D^{th}_L(z,s\&m)=H_0d_L=H_0(1+z)\int^z_0\frac{dz'}{H(z,s\&m)}\end{equation}  The
apparent magnitude is connected to $D_L(z)$ as
\begin{equation}m^{th}(z,s\&m)=M'+5log_{10}(D^{th}_L(z,s\&m)\end{equation} \\
where\begin{equation}M'=M+5log_{10}(\frac{cH^{-1}_0}{Mpc})+25\end{equation}
The best fitting values of the parameters can be obtained through
$\chi^2$ minimization, where
\begin{equation}\chi^2=\sum^{157}_{i=1}\frac{[m^{obs}(z_i)-m^{th}(z_i,s\&m)]^2}{\sigma^2_{m^{obs}(z_i)}}\end{equation}
\\ The corresponding observed $D^{obs}_L(z_i)(i=1,\ldots,157)$
comes from the Gold SN-Ia data set. The observational data are
given as the apparent magnitudes $m^{obs}(z)$ and $1\sigma$ errors
$\sigma_{m^{obs}(z_i)}$. We marginalize the nuisance parameter
$M'$ by defining a new
$\chi^{'2}$: \begin{equation}\chi^{'2}=-2ln\int^{+\infty}_{-\infty}e^{-\chi^2/2}dM'\end{equation}\\
We finally obtain the effective $\chi^2(s\&m)$:
\begin{equation}\chi^{2}(s\&m)=\chi^{2}(M'=0,
s\&m)-\frac{B(s\&m)}{C}\end{equation}\\
where\begin{equation}B(s\&m)=\sum^{157}_{i=1}\frac{[m^{obs}(z_i)-m^{th}(z_i,M'=0,s\&m)]}{\sigma^2_{m^{obs}(z_i)}}\end{equation}
\begin{equation}C=\sum^{157}_{i=1}\frac{1}{\sigma^2_{m^{obs}(z_i)}}\end{equation}
\\Minimizing $\chi^2(s\&m)$ we can find the best fit value of
parameter $s\&m(\chi^2(s_0\&m_0)=\chi^2_{min})$. The $1\sigma$
error on one parameter is determined by the relation $\Delta
\chi^2_{1\sigma}=\chi^2_{s\&m}-\chi^2_{min}=1$, which means the
parameter $s\&m$ in the range $[s_0, s_{1\sigma}]$($[m_0,
m_{1\sigma}]$) with $68\%$ probability. For $2\sigma$
error($95.4\%$ range) $\Delta \chi^2_{2\sigma}=4$ and $3\sigma$
error($99\%$ range) $\Delta \chi^2_{3\sigma}=6.63$.
\begin{center}\vspace{0.5cm}
\includegraphics[width=10cm]{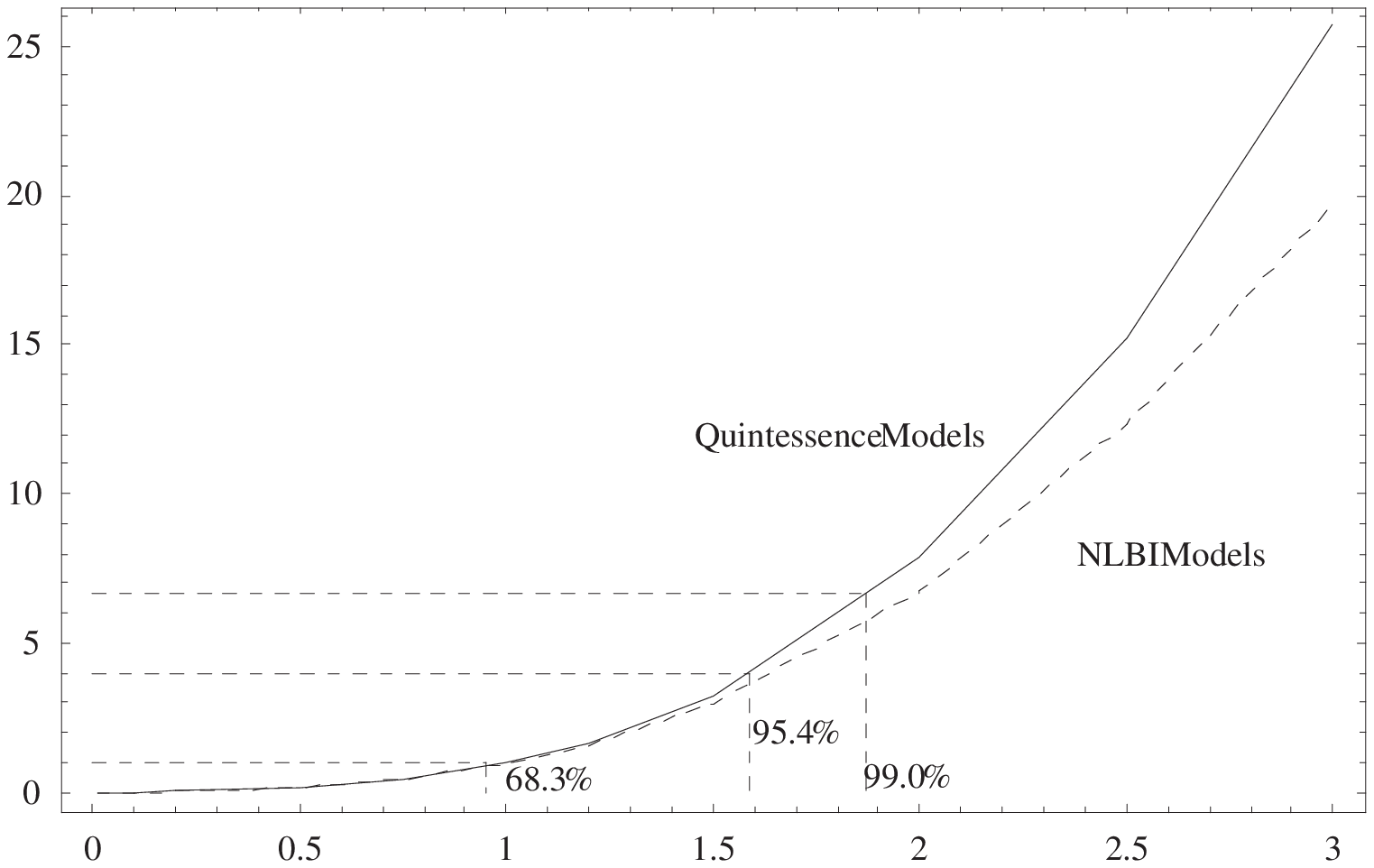} \end{center} \hfill
~\begin{minipage}{5.5in}~Fig9.The differences $\Delta
\chi^2=\chi^2_{s}-\chi^2_{s\simeq0}$ for negative linear potential
$-s\phi$.The horizontal axis represents $s$ and the vertical axis
represents $\Delta \chi^2$. Solid line for quintessence, dot line
for NLBI scalar field. The curve has been sampled at $s=0.01, 0.1,
0.2, 0.5, 0.75, 1.0, 1.2, 1.5, 2.0, 2.5, 3.0$ and the
corresponding points have been joined.
\end{minipage}
\begin{center}\vspace{0.5cm}
\includegraphics[width=10cm]{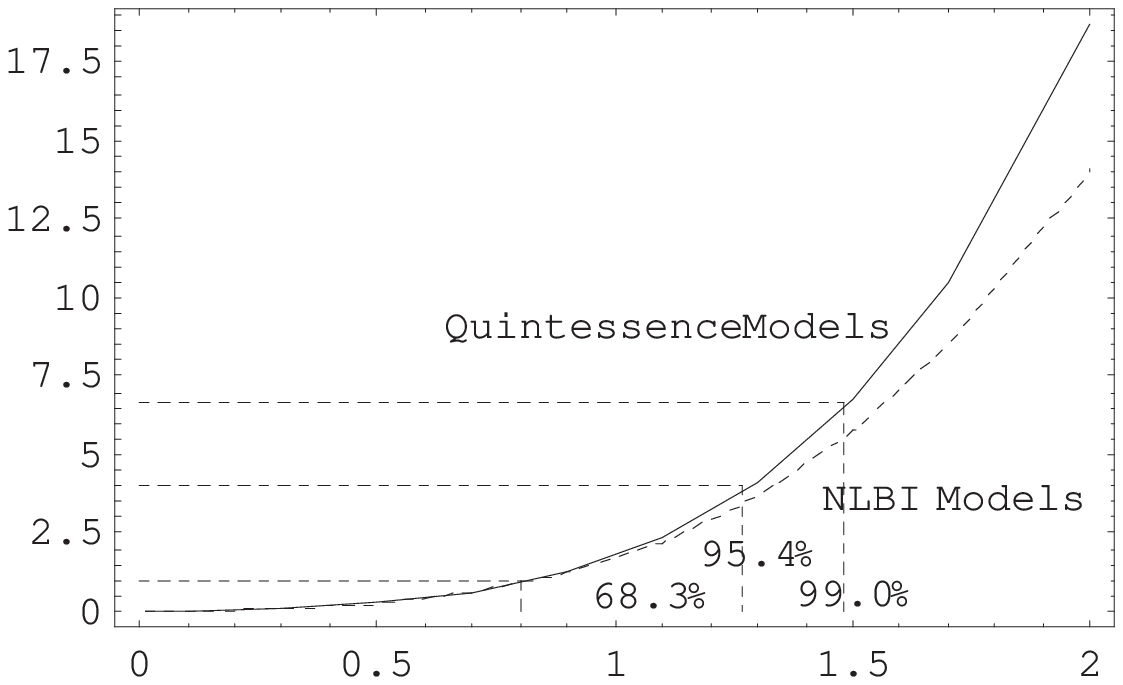} \end{center} \hfill
~\begin{minipage}{5.5in}~Fig10. The differences $\Delta
\chi^2=\chi^2_{m}-\chi^2_{m\simeq0}$ for square potential
$\frac{1}{2}m^2\phi^2$. The horizontal axis represents $m$ and the
vertical axis represents $\Delta \chi^2$. Solid line for
quintessence, dot line for NLBI scalar field. The curve has been
sampled at $m=0.01, 0.1, 0.3, 0.5, 0.7, 0.9, 1.1, 1.3, 1.5, 1.7,
2.0$ and the corresponding points have been jointed.
\end{minipage}

Fig9 and Fig 10 show plots of the difference $\Delta \chi^2$ with
respect to the cosmological constant
($\chi^2(s\simeq0)=\chi^2(m\simeq0)=177.1$). It is implied from
fig9 and fig10 that the $\chi^2$ value of NLBI scalar field is
lower than the value of quintessence for the same $s\&m$ value.
This give us a positive information that NLBI scalar field model
may be superior to quintessence model.
\section{Conclusion and Discussion}
Through the analyses of the two potentials in quintessence model
and NLBI model, we can conclude that:
\par Firstly, the principle we
choose the potential is that the theoretical prediction should be
consistent with the observational universe(such as the observation
of universe age, the CMBR measurement, the SN-Ia observation, the
structure formation and so on). It is to say the predicted
universe need to have nearly same "history"( to account for the
observational data), but the fate of the predicted universe could
have significant differences.
 \par Secondly, for
non-negative potentials, the common feature of the further
universe is that they will continue expanding for ever, though the
fate of the universes with different potentials may be
dramatically altered.
\par Thirdly, if the potential can evolve into negative value,
the universe will evolve continually from expansion($H>0$) to
contraction($H<0$). This is not the case for positive
potential(see fig1 and fig5). The cosmology with negative
potential has been discussed in Ref[32].
\par Finally, comparing our theoretic models with the observational data of SN-Ia, the age of universe $H_0t$,
 the equation of state w and the
transition redshift $z$, we can conclude that the NLBI modes is
consistent with all the observations. Furthermore the result shows
that the NLBI model slightly excels quintessence model. However,
the result also shows that a smaller value of $s\&m$ (a slower
rolling of field $\phi$ correspondingly ) can provide better fits
with observational data, but in this case the difference between
NLBI scalar field and quintessence is not distinct. In order to
get a more convincible result, the effect of NLBI scalar field in
CMB (for instance the late-time ISW effect) should be studied
carefully.
 \hspace*{15 pt}
 \section{Acknowledgement}
 \hspace*{15 pt}We thank L.Perivolaropoulos for providing his mathematical program.
 We thank YunGui Gong for his help in studying the maximum likelihood technique and
marginalization of free parameters at the early stage of this
work. This work is partly supported by National Natural Science
Foundation of China(NNSFC) under Grant No.10573012 and No.10575068
and by Shanghai
 Municipal Science and Technology Commission No.04dz05905. \\

{\noindent\Large \bf References} \small{
\begin{description}
\item {[1]}{P.de Bernardis et al., Nature(London) \textbf{404}:955(2000);\\
            S Hannany et al., Astrophys.J.\textbf{545} L1-L4(2000).}
\item {[2]}{A.G.Riess et al., Astron.J.\textbf{116}:1009(1998);\\
            S.Perlmutter et al., Astrophys.J.\textbf{517}:565(1999);\\
            P.M.Garnavich et al., Astrophys.J. \textbf{509}:74(1998);\\
            J.L.Shievers et al., Astrophys.J.\textbf{591}:590(2003).}
\item {[3]}{W.Freedman et al., Astrophys.J.\textbf{553}:47(2001).}
\item {[4]}{M.Tegmark et al.,(SDSS Collaboration), Phys.Rev.D\textbf{69}:103501(2004);\\
            M.Tegmark et al.,(SDSS Collaboration), Astrophys.J.\textbf{606}:702(2004);\\
            M.Colless et al., astro-ph/0306581.}
\item {[5]}{N.Bahcall, .J.P.Ostriker, S.Perlmutter and P.J.Steinhardt, Science \textbf{284}:1481(1999).}
\item {[6]}{A.Melchiorri, astro-ph/0406652.}
\item {[7]}{S.Weinberg, Rev.Mod.Phys\textbf{61}(1):1(1989);\\
            P.J.E.Peebles, B.Ratra, Rev.Mod.Phys\textbf{75}(2):559(2003);\\
            G.W.Gibbons, hep-th/0302199}
\item {[8]}{P.J.Steinhardt,L.Wang and I.Zlatev, Phys.Rev.D\textbf{59}:123504(1999)\\
            P.Steinhardt, in $Critical Problems in Physics$,
            edited by V.L.Fitch and D.R.Marlow(Princeton University Press, Princeton, NJ(1997)}
\item {[9]}{I.Zlatev, L.Wang and P.J.Steinhardt, Phys.Rev.Lett\textbf{82}:896(1999);\\
            P.Singh, M.Sami and N.Dadhich, Phys.Rev.D\textbf{68}:023522(2003);\\
            J.G.Hao and X.Z.Li, Phys.Rev.D\textbf{67}:107303(2003);\\
            J.G.Hao and X.Z.Li, Phys.Rev.D\textbf{68}:043501(2003);\\
            D.J.Liu and X.Z.Li, Phys.Rev.D\textbf{68}:067301(2003);\\
            T.Chiba, Phys.Rev.D\textbf{66}:063514(2002);\\
            L.Amendola and D.Tocchini-Valentini, Phys.Rev.D\textbf{64}:043509(2001);\\
            X.Z.Li, J.G.Hao and D.J.Liu, Chin.Phys.Lett\textbf{19}:1584(2002);\\
            G.Huey and B.D.Wandelt, astro-ph/0407196;\\
            W.Fang, H.Q.Lu, Z.G.Huang and K.F.Zhang, hep-th/0409080(To appear in Int.J.Mod.Phys.D).}
\item {[10]}{P.J.E.Peebles and B.Ratra, Astrophys.J.Lett\textbf{325}:L17(1988).}
\item {[11]}{C.Wetterich, Nucl.Phys.B\textbf{302}:668(1988);\\
            B.Ratra and P.J.E.Peebles, Phys.Rev.D\textbf{37}:3406(1988);\\
            R.R.Caldwell,R.Dave and P.J.Steinhardt,Phys.Rev.Lett\textbf{80}:1582(1998).}
\item {[12]}{C.Armend\'{a}riz-Pic\'{o}n,T.Damour and V.Mukhanov,Phys.Lett.B\textbf{458}:209(1999);\\
             J.Garriga and V.F.Mukhanov.Phys.Lett.B\textbf{458}:219(1999);\\
             A.G.Riess, astro-ph/0402512; L.P.Chimento,astro-ph/0311613.}
\item {[13]}{C.Armend\'{a}riz-Pic\'{o}n, V.Mukhanov and P.J.Steinhardt,Phys.Rev.Lett\textbf{85}:4438(2000);\\
             C.Armend\'{a}riz-Pic\'{o}n, V.Mukhanov and P.J.Steinhardt,Phys.Rev.D\textbf{63}:103510(2001);\\
             T.Chiba, Phys.Rev.D\textbf{66}:063514;\\
             T.Chiba, T.Okabe and M.Yamaguchi, Phys.Rev.D\textbf{62}:023511(2000);\\
             M.Malquarti, E.J.Copeland, A.R.Liddle and M.Trodden, Phys.Rev.D\textbf{67}:123503(2003);\\
             R.J.Sherrer, Phys.Rev.Lett.\textbf{93)}:011301(2004);\\
             L.P.Chimento, Phys.Rev.D\textbf{69}:123517(2004)}
\item {[14]}{A.Melchiorri, L.Mersini,C.J.Odman and M.Trodden, Phys.Rev.D\textbf{68}:043509(2003).}
\item {[15]}{M.Born and Z.Infeld, Proc.Roy.Soc A\textbf{144}:425(1934).}
\item {[16]}{G.W.Gibbons and C.A.R.Herdeiro, Phys.Rev.D\textbf{63}:064006(2001);\\
             G.W.Gibbons Rev.Mex.Fis.\textbf{49}S1:19-29(2003)(hep-th/0106059);\\
             V.V.Dyadichev, D.V.Gal'tsov and A.G.Zorin,Phys.Rev.D\textbf{65}:084007(2002);\\
             D.N.Vollick, Gen.Rel.Grav.\textbf{35}:1511-1516(2003); gr-qc/0506091 }
\item {[17]}{A.Sen, JHEP\textbf{0207}:065(2002); JHEP\textbf{0204}:048(2002); Mod.Phys.Lett.A\textbf{17}:1797(2002);\\
             G.W.Gibbons, Phys.Lett.B\textbf{537}:1(2002); Glass.Quant.Grav.\textbf{20}:S321-S346(2003).}
\item {[18]}{L.R.Abramo, F.Finelli and T.S.Pereira, Phys.Rev.D\textbf{70}:063517(2004);\\
             J.S.Bagla, H.K.Jassal and T.Padmanabhan, Phys.Rev.D\textbf{67}:063504(2003).}
\item {[19]}{W.Heisenberg, Z.Phys.\textbf{133}:79(1952); \textbf{126}:519(1949); \textbf{113}:61(1939).}
\item {[20]}{H.P.de Oliveira, J.Math.Phys.\textbf{36}:2988(1995).}
\item {[21]}{H.Q.Lu, Int.J.Modern.Phys.D\textbf{14}:355(2005); H.Q.Lu et al. hep-th/0504038.}
\item {[22]}{T.Taniuti,Prog.Theor.Phys.(kyoto) Suppl\textbf{9}:69(1958).}
\item {[23]}{L.Perivolaropoulos, Phys.Rev.D\textbf{71}:063503(2005).}
\item {[24]}{J.Garriga and A.Vilenkin, Phys.Rev.D\textbf{61}:083502(2000); Phys.Rev.D\textbf{67}:043503(2003);\\
             J.Garriga A.Linde and A.Vilenkin, Phys.Rev.D\textbf{69}:063521(2004);\\
             P.P.Avelina, Phys.Lett.B\textbf{611}:15(2005).}
\item {[25]}{A.D.Linde, Phys.Lett.B\textbf{129}:177(1983).}
\item {[26]}{A.Kamenshchik,U.Moschella and V.Pasquier,Phys.Lett.B\textbf{511}:265-268(2001);\\
             N.Bilic, G.B.Tupper and R.D.Viollier,Phys.Lett.B\textbf{535}:17-21(2002);\\
             M.C.Bento, O.Bertolami and A.A.Sen, Phys.Rev.D\textbf{70}:083519(2004);\\
             M.C.Bento, O.Bertolami and A.A.Sen, Phys.Rev.D\textbf{66}:043507,(2002);\\
             M.C.Bento, O.Bertolami and A.A.Sen, Gen.Rel.Grav.\textbf{35}:2063-2069(2003);\\
             M.C.Bento, O.Bertolami and A.A.Sen, Phys.Lett.B\textbf{575}:172-180,(2003);\\
             M.C.Bento, O.Bertolami and A.A.Sen, Phys.Rev.D\textbf{67}:063003(2003);\\
             P.T.Silva and O.Bertolami, Astrophys.J.\textbf{599}:829-838(2003);\\
             O.Bertolami, A.A.Sen, S.Sen and P.T.Silva, Mon.Not.Roy.Astron.Soc.\textbf{353}:329(2004);\\
             M.C.Bento, O.Bertolami, N.M.C.Santos and A.A. Sen, Phys. Rev.D\textbf{71}:063501(2005);\\
             N.Bilic, G.B.Tupper and R.D.Viollier, astro-ph/0207423;\\
             J.C.Fabris, S.V.B.Goncalves and P.E.de Souza, astro-ph/0207430;\\
             A.Dev, J. S.Alcaniz and D.Jain, Phys.Rev.D\textbf{67}:023515(2003);\\
             V.Gorini, A.Kamenshchik and U.Moschella, Phys.Rev.D\textbf{67}:063509(2003);\\
             M.Makler, S.Q.de Oliveira and I.Waga, Phys.Lett.B\textbf{555}:1(2003);\\
             R.Colistete Jr., J.C.Fabris, S.V.B.Goncalves and P.E.de Souza, gr-qc/0210079;\\
             D.j.Liu and X.Z.Li, Chin.Phys.Lett.\textbf{22}:1600-1603(2005);\\
             X.Zhang, F.Q.Wu and J.F.Zhang,astro-ph/0411221;\\
             Y.G.Gong, JCAP0503:007(2005);\\
             F.Perrotta, S.Matarrese and M.Torki, Phys.Rev.D\textbf{70}: 121304(2004);\\
             V.Gorini, A.Kamenshchik, U.Moschella and V.Pasquier, gr-qc/0403062;\\
             J.S.Alcaniz and J.A.S.Lima, Astrophys.J.\textbf{618}:16(2005);\\
             N.Bilic, R.J.Lindebaum, G.B.Tupper and R.D.Viollier, JCAP0411:008(2004);\\
             J.V.Cunha, J.S.Alcaniz and J.A. S.Lima, Phys.Rev.D\textbf{69}:083501(2004).}
\item {[27]}{J.K.Erickson, R.R.Caldwell, P.J.Steinhardt, C.Armend\'{a}riz-Pic\'{o}n and V.Mukhanov,
                        Phys.Rev.Lett\textbf{88}:121301(2002);\\
              S.Dedeo, R.R.Caldwell and P.J.Steinhardt, Phys.Rev.D\textbf{67}:103509(2003).}
\item {[28]}{R.Bean and O.Dore, Phys.Rev.D\textbf{69}:083503(2004).}
\item {[29]}{http:// leandros.physics.uoi.gr/negpot.htm.}
\item {[30]}{E.Carretta et al., Astrophys.J\textbf{533}:215(2000);\\
             W.L.Freedman et al., Astrophys.J\textbf{553}:47(2001);\\
             H.B.Richer et al., Astrophys.J\textbf{574}:L15(2002);\\
             B.M.S.Hansen et al., Astrophys.J\textbf{574}:L155(2002);\\
             L.Krauss, Astrophys.J\textbf{596}:L1(2003);
             J.Cepa,Astron.Astrophys.\textbf{422}:831(2004).}
\item {[31]}{A.G.Riess, Astrophys.J\textbf{560}:49(2001);\\
             M.S.Turner and A.G.Riess, Astrophys.J\textbf{569}:18(2002);\\
             R.A.Daly and S.G.Djorgovski, Astrophys.J\textbf{597}:9(2003);\\
             Y.G.Gong and C.K.Duan, Class.Quantum.Grav.\textbf{21}:3655(2004).}
\item {[32]}{G.Felder, A.Frolov, L.Kofman and A.Linde, Phys.Rev.D\textbf{66}:023507(2002).}
\end{description}}
\end{document}